\documentclass[twocolumn,aps,prl,superscriptaddress,longbibliography]{revtex4-2}
\usepackage{graphicx}
\usepackage{color}
\usepackage{amsmath,amsfonts,amssymb}
\usepackage{comment}
\usepackage[svgnames]{xcolor}
\usepackage [latin1]{inputenc}
\usepackage[colorlinks,linkcolor=MediumBlue,citecolor=MediumBlue,urlcolor = MediumBlue]{hyperref}
\usepackage{url}

\newcommand{\p}{\mathbf{p}}
\newcommand{\Q}{\mathbf{Q}}
\newcommand{\br}{\mathbf{r}}
\newcommand{\n}{\hat{\mathbf{n}}}

\newcommand{\brho}{\bar{\rho}}

\newcommand{\bS}{\bar{S}}
\newcommand{\bPhi}{\bar{\Phi}}

\begin{document}

\title{Density-protected states in active matter under virtual confinement}

\author{Giuseppe Fava}
\email{gfava.phd@gmail.com}
\affiliation{Center for Soft Condensed Matter Physics and Interdisciplinary Research \& School of Physical Science and Technology, Soochow University, Suzhou 215006, China}
\affiliation{Laboratoire de Physique Th{\'e}orique de la Mati{\`e}re Condens{\'e}e (LPTMC), Sorbonne Universit{\'e}, Paris 75005, France}

\author{Francesco Ginelli}
\email{francesco.ginelli@uninsubria.it}
\affiliation{Dipartimento di Scienza e Alta Tecnologia and Center for Nonlinear and Complex Systems, Universit{\`a} degli Studi dell'Insubria, Como, Italy}
\affiliation{INFN sezione di Milano, Milano, Italy}

\author{Beno\^{i}t Mahault}
\email{benoit.mahault@umontpellier.fr}
\affiliation{Laboratoire Charles Coulomb (L2C), UMR 5221 CNRS--Universit{\'e} de Montpellier, Montpellier F-34095, France}
\affiliation{Max Planck Institute for Dynamics and Self-Organization (MPI-DS), G\"ottingen, Germany}

\date{\today}

\begin{abstract}
We investigate photo-responsive structure formation in a minimal model
of dry active nematics. Combining microscopic simulations with the
analysis of the corresponding hydrodynamic theory, we show that the
system generically self-assembles into a dense, nematically ordered
ring at the boundary of compact illumination patterns. Remarkably,
this boundary structure gives rise to a disordered
core whose density is self-selected and independent of the global particle density.
Our analysis reveals that these protected states emerge from a generic interplay
between local nematic alignment and curvature-driven active currents.
These results identify a robust route to boundary-induced
structure formation in active matter
and provide experimentally testable predictions.
\end{abstract}

\maketitle

Active filaments, which take forms as diverse as cytoskeleton components propelled by molecular motors, bacteria, or worms, are prominent examples of active matter~\cite{WinklerJCP2020}. 
Continuously dissipating energy into directed motion, 
these active units are able to spontaneously self-organize into collectively moving assemblies or intricate patterns~\cite{NdlecNature1997,SchallerNature2010,SuminoNature2012,SugiNatCom2019,FaluwekiPRL2023,KurjahnNC2024,CammannComPhys2024}. 
Beyond the insights it provides into understanding the emergence of self-organization in the living world, 
active matter also presents appealing avenues for engineering biomimetic materials with programmable properties~\cite{BowickPRX2022}.

Given the diversity of phenomena it encompasses, the design of reliable strategies to control active matter towards robust states poses a significant challenge, 
and is currently attracting growing interest~\cite{NortonPRL2020,LiuNature2021,ShankarNRP2022,ShankarPNAS2024,DavisPRX2024}. 
Theoretical and experimental investigations have shown that active systems are indeed
particularly sensitive to mechanical confinement~\cite{BenDorPRE2022,CodinaPRL2022,Lenzini2024,Fava2024PRL} or patterning~\cite{KaiyrbekovPNAS2023,ZhaoPRL2024,ZhaoNatCom2025}, with effects
  depending strongly on geometry and interactions with boundaries.
Unfortunately, dynamical control strategies relying on mechanical confinement are difficult to implement 
at the micro-scales characteristics of biological agents.

However, the fact that active assemblies are often composed of agents able to detect and respond to environmental cues 
offers alternative strategies that bypass the difficulties associated with hard boundaries.
For example, a large number of microorganisms exhibit photo-responsive behavior, 
allowing them to spontaneously migrate to regions with favorable light conditions~\cite{WildeFEMS2017}. 
This property also extends to motor protein-biofilament assemblies~\cite{SchupplerNatCom2016,RossNature2019,ZhangNatMat2021}, genetically engineered bacteria~\cite{ArltNatCom2018,FrangipaneeLife2018,Massana-CidNatCom2022}, active colloids~\cite{LozanoNatCom2016,SokerPRL2021,AlvarezNatCom2021,vanKesterenPNAS2023}, or light-sensitive robots~\cite{MijalkovPRX2016}
whose activity can be spatially structured by illumination patterns. 
As a result, light patterns can guide these systems toward specific locations and act as a form of {\it virtual confinement} without imposing external mechanical forces.
When the particles are elongated, such aggregation is further accompanied by the emergence of orientational order, 
which drives the formation of more complex structures, including asters~\cite{RossNature2019}, 
topological defects~\cite{ZhangNatMat2021}, or ring-like vortices~\cite{KurjahnNC2024}.
This phenomenon thus enables the engineering of tunable material architectures.

The principles underlying such structure formation remain incompletely understood. 
Inspired by recent observations of light-responsive filamentous cyanobacteria~\cite{KurjahnNC2024},
we investigate pattern formation in a minimal model of
self-propelled agents combining nematic alignment with velocity reversals triggered by local illumination gradients.
Realizing virtual confinement via an illuminated shape, we show that particles spontaneously align and accumulate at the light-dark interface,
forming patterns similar to that reported in experiments.
Over a broad region of the phase diagram and independently of the confining geometry, we show that these boundary
structures screen the interior of the illuminated domain, producing a
disordered {\it protected core} characterized by a selected local
density independent of the overall density of the system. 
By mapping our model onto a hydrodynamic-level description, 
we show that these protected states arise from the interplay between local alignment and active currents driven by nematic order curvature.
This uncovers a generic mechanism for boundary-induced structure formation in active matter and enabling experimentally testable predictions.

In contrast to previous works~\cite{KurjahnNC2024,FaluwekiPRL2023,CammannComPhys2024}, 
we model filamentous bacteria by considering point-like particles moving in two dimensions with constant speed $v_0$. 
To account for the head-tail symmetry of the bacterial body, particles align their unit self-propulsion direction $\hat{\mathbf{n}}$ with a nematic symmetry~\cite{GinelliPRL2010,Romanzuck2024,PatelliPRL2019}. 
In addition, we model pairwise repulsion between the particles by
means of a repelling torque with characteristic strength $\beta$.
The position $\br_i^t$ and orientation $\n_i^t = (\cos\theta_i^t, \sin\theta_i^t)^T$ of the $i^{\rm th}$ 
particle evolve in discrete time $t$ as
\begin{subequations}
	\label{micro}
	\begin{align}
		\label{r_i}
		\br_i^{t+1} & = \br_i^t + v_0 \n_i^{t+1}, \\
		\label{theta_i}
		\n_i^{t+1} & = %\n_i^{t} +
		({\cal R}_\eta \circ \vartheta) 
		\left[\left\langle \left({\rm sign}[\n_i^{t}\cdot \n_j^{t}] \n_j^t 
		+\beta \hat{\br}_{ji}^t \right) \right\rangle_{j \sim i}\right],
	\end{align}
\end{subequations}
where $\vartheta(\mathbf{u}) = \mathbf{u} / |\mathbf{u}|$,
while the operator ${\cal R}_\eta$ performs uncorrelated random rotations with an 
angle uniformly drawn in $(-\eta \pi, \eta \pi]$.
The average $\langle \cdot \rangle_{j\sim i}$ runs over all particles $j$ 
within unit distance of $i$, including $i$ itself. 
The ${\rm sign}$ operator in the first term on the r.h.s.\ of Eq.~\eqref{theta_i} 
ensures that the alignment is nematic, 
as it leads particle orientations forming an acute angle to align, and anti-align otherwise.
In the second term, $\hat{\br}_{ji}^t = \vartheta(\br_i^t - \br_j^t)$ is the unit vector pointing from particle $j$ to $i$ (we use $\hat{\br}_{ii}^t = \mathbf{0}$),
such that for $\beta > 0$ this term leads particles to reorient away from their neighbors.
Note that this choice of repelling torque does not prevent positional overlaps of nearby self-propelled particles across their effective extension, consistently with the quasi-2D nature of typical experimental realizations~\cite{SchallerNature2010,SuminoNature2012,FaluwekiPRL2023,KurjahnNC2024}. 
Quasi two-dimensional motion also justifies the choice of a dry model, neglecting momentum transfer to the surrounding fluid due to boundary dissipation~\cite{DADAM}.

%%%%%%%%%%%%%%%%%%%%%%%%%%%%%%
\begin{figure}[t!]
\includegraphics[width=\linewidth]{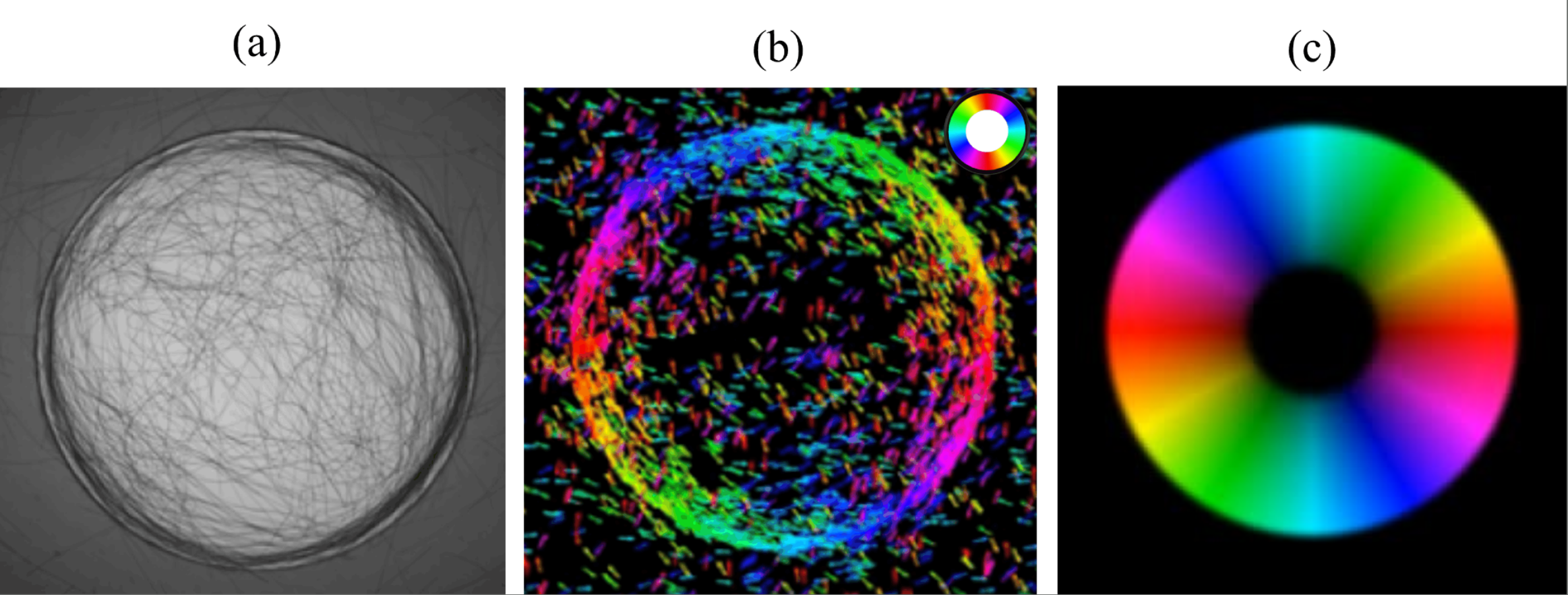}
\caption{(a) Experimental ring configuration formed by {\it O.lutea}
  cyanobacteria at the edge of an illuminated region. Reprinted from
  \cite{KurjahnNC2024}. (b) Typical snapshot of a stationary ring
  configuration obtained from the microscopic dynamics
    (\ref{micro}). The color-disk (top right) measures the
  local nematic angle $\Phi$. Black indicates the absence of
  particles. Here $\beta=0.3$, $\rho_0 = 0.5$, $L=512$, $R=100$. 
  (c) Stationary solution obtained from continuum
  Eqs.~(\ref{eq_hydro}) with $\rho_0 = 1$, $L = 128$, $R = 32$ and other parameters as in Fig.~\ref{hydro}.}
\label{sketch_interactions}
\end{figure}
%%%%%%%%%%%%%%%%%%%%%%%%%%%%%%

We implement virtual confinement by introducing an illuminated disk of radius $R$
and mimic the photo-responsive behaviour displayed by cyanobacteria~\cite{RisserAnnrev2025} 
or other photo-sensitive microorganisms~\cite{WildeFEMS2017}:
whenever a particle $i$ crosses the disk border while coming from the bright area, 
it experiences a reversal of its polarity, $\n_i^{t+1} \to - \n_i^{t+1}$.
On the other hand, particles entering the disk from the dark surroundings do not revert.
Naturally, such biased reversals trap particles inside of the illuminated disk,
reflecting the active accumulation of photo-responsive organisms in regions with favorable light conditions.

All simulations of Eqs.~\eqref{micro} were performed with $N$ particles in periodic square domains of linear size $L > R$.
For simplicity, we only consider two control parameters, the repulsion strength $\beta$ and the mean
particle density $\rho_0=N/L^2$, fixing $v_0 = 0.3$ and $\eta = 0.1$,
but we have verified the robustness of our results to changes of these parameters.
In the absence of virtual confinement (i.e., with homogeneous illumination) 
the phase diagram of this model has been worked out
in Ref.~\cite{PatelliPRL2019}: depending on $\rho_0$ and $\beta$, 
the dynamics exhibits homogeneous disordered or nematically ordered states, 
as well as a chaotic phase populated by $\pm\tfrac{1}{2}$charged topological defects.

Once virtual confinement is introduced, the model~\eqref{micro} 
reproduces, for generic parameters, the typical phenomenology observed
in experiments [see Fig.~\ref{sketch_interactions}(a)], with the emergence of a high-density ring of particles aligned with the illuminated disk interface, 
as shown in Fig.~\ref{sketch_interactions}(b). 
We quantify local alignment with the instantaneous symmetric and
traceless nematic tensor
${\Q}({\bf r},t) = \sum_{i \in V({\bf r})}[ {\n}_i^t \otimes {\n}_i^t - {\mathbf I}/2]$,
where the sum runs over all particles inside a square box of unit size centered on position ${\bf r}$ and $\otimes$ is the outer product. 
The strength of ${\Q}$ is given by the nematic amplitude $S({\bf r},t) = 2\sqrt{Q_{11}({\bf r},t)^2+Q_{12}({\bf r},t)^2}$,
while its orientation is captured by the nematic angle $\Phi({\bf r},t)= \tfrac{1}{2} \arg[Q_{11}({\bf r},t) + iQ_{12}({\bf r},t)]$.
In stationary state, we consider the time-averaged nematic field $\bar{\Q}({\bf r})=\langle {\Q}({\bf r},t)\rangle_t$ from which we evaluate the time-averaged amplitude $\bar{S}({\bf r})$ and orientation $\bar{\Phi}({\bf r})$.

\begin{figure}[t]
	\includegraphics[width=1\linewidth]{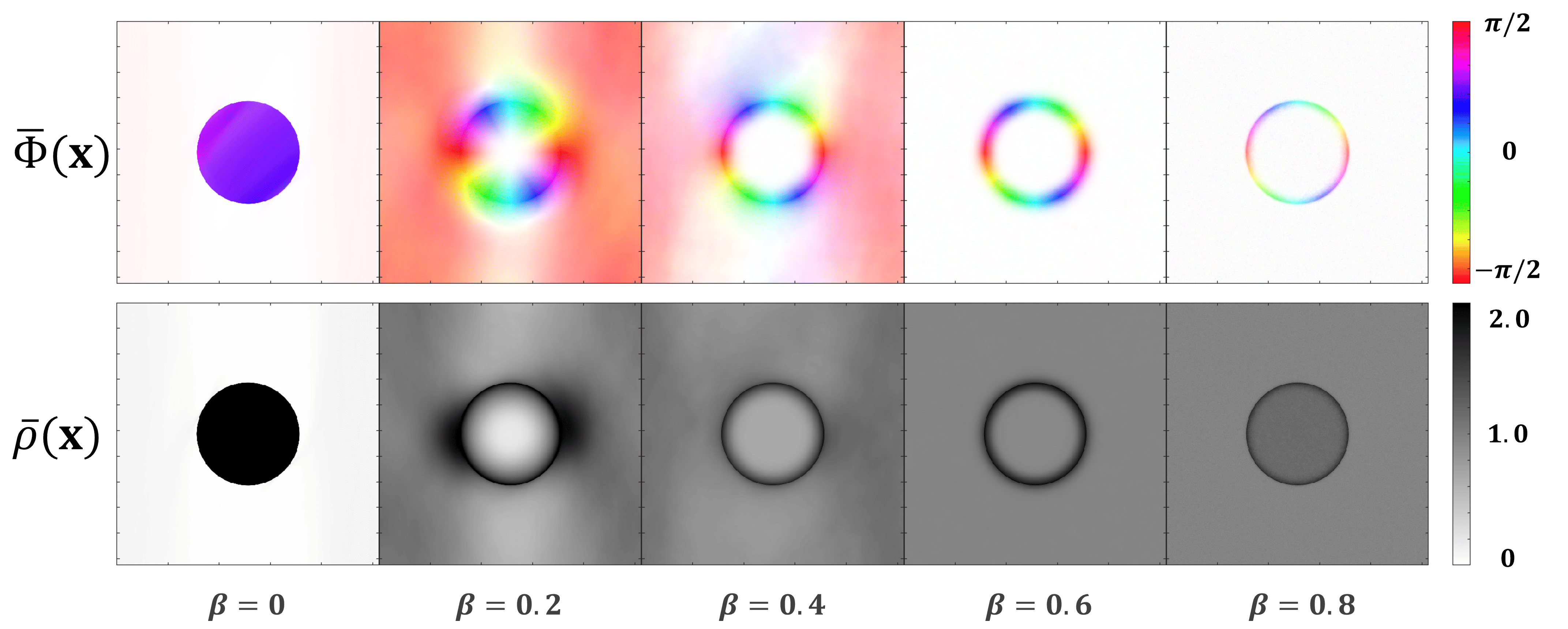}
	 \caption{Stationary configurations at fixed global density $\rho_0 = 1$ 
	 and for $\beta\in[0,0.8]$. 
 	The top row shows the (color-coded) time-averaged nematic angle field, 
	where disordered regions appear white.
	The bottom row shows the time-averaged density (gray-scale).
 	System size: $L = 512$ and $R = 100$. 
        }
	\label{phase_diagram}
\end{figure}

Figure~\ref{phase_diagram} shows that ring structures emerge for
$\beta > 0$ regardless of the phase of the system outside the
illuminated disk, highlighting the robustness of this self-organized
state to changes in the background far field dynamics. 
Rings always form provided that $R$ is sufficiently large and the global particle density $\rho_0$
is sufficiently high for nematic order to develop inside the disk. The
main requirement for ring formation is the presence of repelling
torques (see \hyperlink{appA}{Appendix A} in End Matter).
In their absence, the virtual confinement mechanism is almost perfectly trapping~\footnote{At least
for the relatively small noise amplitudes considered in this work.},
such that nearly all particles accumulate within the illuminated disk, 
where they form a homogeneously ordered domain (leftmost panel of
Fig.~\ref{phase_diagram}). Pairwise repelling torques combined with strong density gradients at the boundary, on the other
hand, allow particles to escape the illuminated region, resulting in a
nonvanishing outside density.

%%%%%%%%%%%%%%%%%%%%%%%%%%
%%%%%%%%%%%%%%%%%%%%%%%%%%

\begin{figure}[t!]
	\includegraphics[width=1.\linewidth]{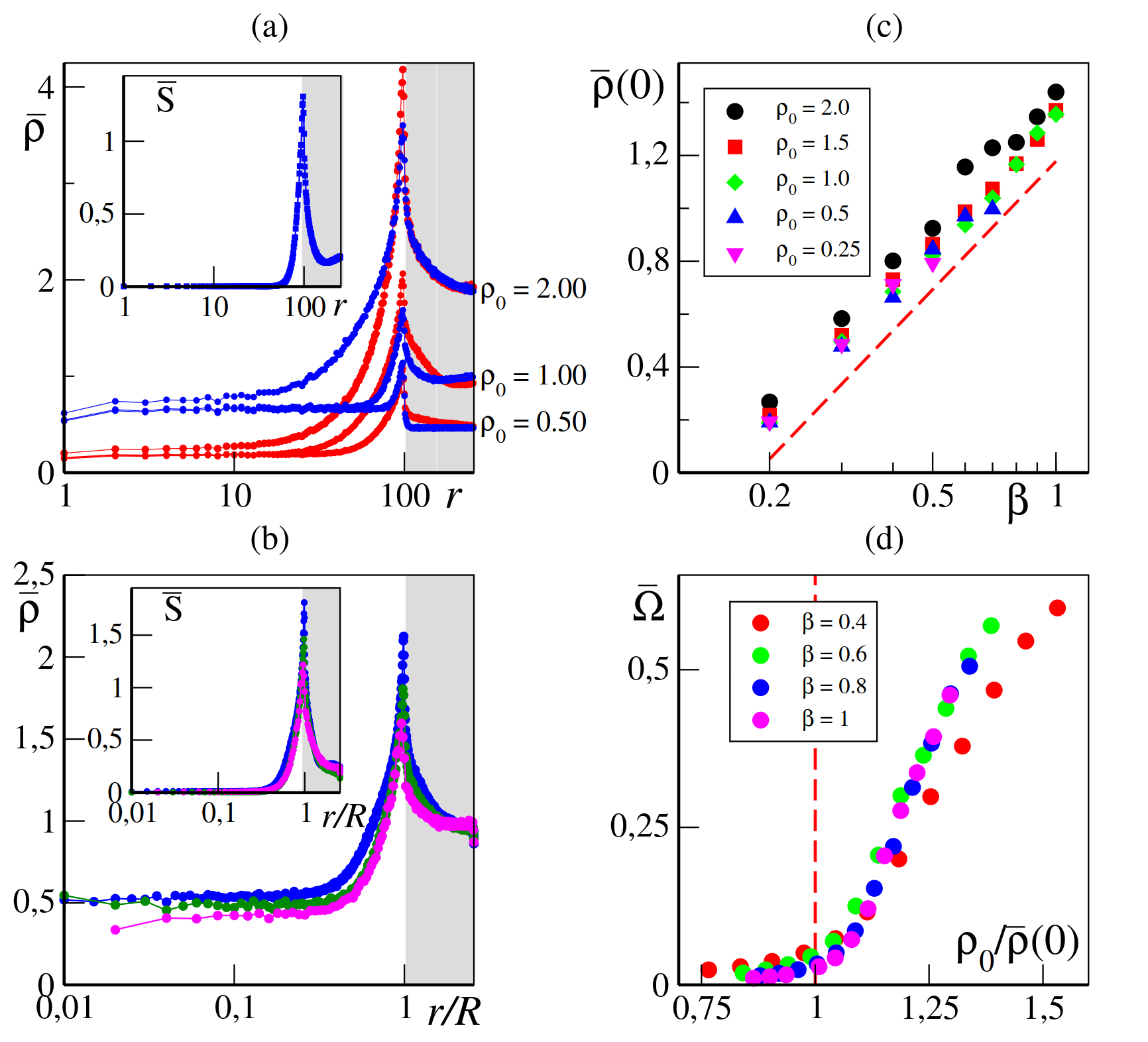}
	 \caption{(a) Radially averaged stationary density profiles 
	 for $\beta=0.2$ (red) and $0.4$ (blue) and $\rho_0 = 0.5$, $1$, $2$.
	 The curves corresponding to equal $\rho_0$ 
	 overlap for $r > R$ (gray region).
	 Inset: nematic amplitude for $\rho_0 = 1$.
	 (b) Stationary profiles for fixed $\beta = 0.3$ and $\rho_0 =
         1$ as functions of $r/R$
	 with system size $L = 256$ (magenta), $L = 512$ (green) 
	 and $L = 1024$ (blue) keeping $L/R = 5.12$.
         (c) Core density $\bar{\rho}(r=0)$ as a function of $\beta$ 
	 for different values of $\rho_0$. 
	 The red dashed line marks the best logarithmic fit 
	 (vertically shifted for clarity) $\bar{\rho}(0)=0.7 \ln(\beta/0.15)$
	 (d) Nematic order parameter measured in homogeneous systems of linear size $L=100$ 
	 as a function of $\rho_0$ 
	 rescaled by the core density for different values of $\beta$.
	 The vertical dashed line marks $\rho_0/\bar{\rho}(0)=1$.}
 \label{profiles}
\end{figure}

We now quantitatively characterize the ring configuration and discuss density protection in the interior.
We report in Fig.~\ref{profiles}(a) radially averaged profiles of the
stationary coarse-grained density $\bar\rho(r)$ and nematic amplitude $\bar S(r)$ for several values of $\beta$ and $\rho_0$.
While local particle accumulation at the disk boundary is
clearly marked by peaks at $r = |{\bf r}| \lesssim R$, 
the value of $\bar\rho(r)$ for $r\to0$ 
{\it is independent of the global density}.
Increasing $\rho_0$, the density in the outer region and near the boundary grows linearly with $\rho_0$,
while its value at the centre of the disk remains practically constant and only depends on the repulsion strength $\beta$.
This feature is confirmed in Fig.~\ref{profiles}(c),
which shows that the core density $\bar\rho(r = 0)$ 
for $\rho_0$ covering roughly an order of magnitude
collapses onto a master curve that exhibits a logarithmic growth 
with the repulsion strength $\beta$. 
The presence of the ring shields the inner core area, 
making it a protected configuration with a selected local density and vanishing nematic order (inset of Fig.~\ref{profiles}(a)).
While the ring width varies non-trivially with $\rho_0$ and $\beta$, 
we show in Fig.~\ref{profiles}(b) that it grows linearly with $R$,
provided that the ratio $R/L$ stays fixed.
Hence, the ring is an extensive structure and its properties survive in the thermodynamic limit.

%%%%%%%%%%%%%%%%%%%%%%%%%%
%%%%%%%%%%%%%%%%%%%%%%%%%% 

We next discuss the mechanism underlying density protection in the ring core.
In homogeneous space, the dynamics~\eqref{micro} is characterized by a threshold density $\rho_c(\beta)$ 
beyond which global nematic order emerges.
Performing simulations without the light pattern,
we obtain curves such as those of Fig.~\ref{profiles}(d) showing that 
the global nematic order parameter $\bar\Omega = \langle|\langle e^{2 i \theta_k^t} \rangle_k|\rangle_t$, 
vanishes for $\rho_0  < \rho_c(\beta)$ and exhibits a sharp increase for $\rho_0 \gtrsim \rho_c(\beta)$. 
Interestingly, rescaling $\rho_0$ with the core density $\bar{\rho}(0)$ obtained for the same value of $\beta$,
the curves of Fig.~\ref{profiles}(d) collapse in the transition region around $\rho_0 = \bar{\rho}(0)$.
This implies that $\bar{\rho}(0) \approx \rho_c(\beta)$,
suggesting that the system self-organizes through active nematic fluxes~\cite{RamaswamyEPL2003,Ramaswamy} that displace particles radially from the core, up to the point at which local nematic order is suppressed.
At the boundary, nematic order persists but active fluxes are compensated by the light-induced trapping that confines particles inside.
As a result, a stationary ring configuration with protected core emerges.
This mechanism thus holds for generic compact illumination patterns, such that we obtain similar density-protected states in both convex and nonconvex geometries (\hyperlink{appB}{Appendix B}).

The generality, robustness, and extensivity of the ring structure call for a continuum description.
Here, we keep a minimal approach by considering a reduced set of field equations describing the microscopic dynamics~\eqref{micro}.
In addition to the density $\rho({\bf r},t)$ and nematic $\Q({\bf
  r},t)$ fields, the polar symmetry of the self-propulsion dynamics 
  requires to consider the coarse-grained polar field ${\bf p}({\bf
  r},t) = \sum_{i \in V({\bf r}) }\n_i^t$ advecting the density. 
The simplest set of equations capturing the ring formation reads
\begin{subequations} \label{eq_hydro}
    \begin{align} 
        \label{eq_hydro_rho}
        & \partial_t \rho + \nabla \cdot {\bf J} = 0, \\
        \label{eq_hydro_p}
        & \partial_t \p = -\alpha \p -\pi_0 \nabla \rho - \pi_2 \nabla \cdot \Q  
        - \kappa g(r) \rho \n(\varphi), \\
        \label{eq_hydro_Q}
        & \partial_t \Q = \left[\mu(\rho\!-\!\rho_c) \!-\! \xi {\rm Tr}(\Q^2)\right]\Q  
       \! +\! D_Q \nabla^2 \Q \!-\! \pi_1[\nabla \otimes \p]_{\rm ST}\,,
    \end{align}
  \end{subequations}
where ${\bf J}=v_0{\bf p} - D_\rho \nabla \rho$ is the density
  flux. 
  Unlike systematically coarse-grained theories, which typically present many nonlinearities~\cite{peshkov2012nonlinear,MahaultJSTAT2018,BertinPRE2015} making analytical investigations challenging,
this minimal truncation retains only the lowest-order symmetry-allowed
terms needed to reproduce local nematic ordering and
curvature-induced active nematic currents~\cite{RamaswamyEPL2003}.
For simplicity, we focus on the circular geometry so that polarity reversals at the light-dark interface are modeled through an external field $\kappa g(r) \n(\varphi)$, 
where $\n(\varphi)$ is the radial unit vector in polar coordinates ($r,\varphi$)
and $g(r)$ is a normalized function sharply peaked at $r = R$ 
and vanishing when $r \ll R$ or $r \gg R$. 
Concretely, reversals locally bias the polarity towards the
illuminated disk center. Within our minimal continuum description, the
effect of microscopic repelling torques is captured
phenomenologically by the effective reversal rate $\kappa$, which
controls the balance between trapping and leakage across the illuminated region boundary.
Polar field dynamics also includes a damping term ($\propto \alpha>0$),
suppressing polar order far from the interface, as well as isotropic
($\propto \pi_0$) and anisotropic ($\propto \pi_2$) pressure terms.
Equation~\eqref{eq_hydro_Q} features a Landau potential with
a density-dependent linear term that accounts for local nematic
ordering when $\rho>\rho_c$, and a ``polarity (flow) aligning term''~\cite{DeGennesProst}
($\propto \pi_1$), where $[\cdot]_{\rm ST}$ denotes the symmetric traceless part.
Importantly, in our dry model the transport terms with positive couplings $v_0$ and $\pi_{0,1,2}$ 
encode the active nature of the dynamics, 
as they directly result from microscopic self-propulsion~\cite{peshkov2012nonlinear}.

\begin{figure}[t!]
	\includegraphics[width=1.\linewidth]{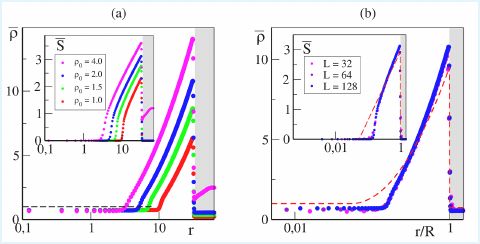}
	 \caption{Stationary solutions of the continuum model~\eqref{eq_hydro}.
	 Radial density and nematic order (inset) profiles for 
	 different values of the global density $\rho_0$(a) and
     	 three different system sizes $L$(b). 
	 In (a) $L = 128$, $R = 32$ and the black dashed line indicates $\rho_c = 1$.
	 In (b), $\rho_0 = 2$, $R = \tfrac{1}{4}L$, 
	 and the red dashed line shows the approximated analytical solution (see text).
	 Parameters: $v_0 = \alpha = 1$, $\pi_0 = \pi_1 = \pi_2 = \kappa = 0.5$, $\bar{\mu} = \xi = 1$, $D_Q = D_\rho = 0.01$. 
	 In both panels, the gray background indicates the outside (dark) region.}
 \label{hydro}
\end{figure}

As shown in Figs.~\ref{sketch_interactions}(c) and~\ref{hydro}, 
direct numerical simulations of Eqs.~\eqref{eq_hydro} (see~\hyperlink{appE}{Appendix E} for details)
lead to solutions that recapitulate the essential characteristics of the ring.
The steady-state density and nematic amplitude are radially symmetric 
and peaked at $r \lesssim R$, 
leaving a uniform disordered region around the core of the illuminated disk.
In agreement with microscopic simulations, the core density $\bar{\rho}(0)$ is protected, 
as it does not vary with the average density $\rho_0$ [Fig.~\ref{hydro}(a)], 
contrary to the ring density, $\rho_R$, which grows with $\rho_0$.
In addition, the ring width increases linearly with $R$ [Fig.~\ref{hydro}(b)], 
confirming that the structure is extensive. 

%%%%%%%%%%%%%%%%%%%%%%%%%%
%%%%%%%%%%%%%%%%%%%%%%%%%% 

To understand how these features emerge, 
we look for stationary solutions of Eqs.~\eqref{eq_hydro}.
Setting time derivatives to zero, we express the polar order as
$
\bar\p = -\alpha^{-1}[\pi_0 \nabla \bar\rho + \pi_2 \nabla \cdot \bar\Q  
        + \kappa g(r) \bar\rho \n(\varphi)]
$.
Using the symmetry of the ring structure,
we seek inhomogeneous solutions of the form 
$\rho \!\!=\!\!\bar\rho(r)$,
$Q_{11}\!\!=\!\!\tfrac{1}{2}\bar S(r)\cos [2 \bar\Phi(\varphi)]$ and $Q_{12}\!\!=\!\!
\tfrac{1}{2}\bar S(r)\sin [2 \bar\Phi(\varphi)]$.
Since nematic rings are associated with a vanishing net density flux, we look for solutions satisfying ${\bf J}=\bf 0$~\footnote{Nonvanishing divergence-free radial currents would imply an unphysical particle source or sink at the illuminated disk center.}. 
Replacing our ansatz and the expression of $\bar\p$ in Eq.~\eqref{eq_hydro_rho}, we obtain (in complex notations for compactness):
\begin{equation}
\label{ss_density}
\tilde{D}_\rho \bar{\rho}' + \frac{\kappa v_0}{\alpha} \bar{\rho} g 
+ \tilde{\pi}_2 e^{i 2(\bar{\Phi}-\varphi)}\left(\bar{S}' +
                                   \frac{2}{r} \bar{S} \bar{\Phi}'
                                     \right) = 0,     
\end{equation}
where $\tilde{D}_\rho = D_\rho + v_0\pi_0/\alpha$ 
and $\tilde{\pi}_2 = \tfrac{1}{2}v_0\pi_2/\alpha$,
while primes denote derivatives.
As detailed in~\hyperlink{appC}{Appendix C}, 
Equation~\eqref{ss_density} admits three types of solution.
When nematic order is homogeneous ($\bar{S}' = \bar{\Phi}' = 0$),
it predicts nearly uniform density profiles inside and outside
the disk, $\bar{\rho}(r) = \bar{\rho}(0)\exp[-\zeta \int_0^r dx \, g(x)]$,
where the dimensionless parameter $\zeta = \kappa v_0 / (\alpha \tilde{D}_\rho)$
determines the relative density ratio.
For $g(r)$ sharply peaked at $r = R$, we approximate $g(r) \approx \delta(r - R)$,
such that
\begin{align} \label{eq_rho_noring}
\bar{\rho}(r \le R) = \bar{\rho}(0)~, & &
\bar{\rho}(r > R) = \bar{\rho}(0)e^{-\zeta}~,
\end{align}
with $\bar{\rho}(0) \propto \rho_0 $ determined by density normalization.

When nematic order is inhomogeneous,
taking the imaginary part of Eq.~\eqref{ss_density} imposes that $\bPhi(\varphi) = \varphi + k\tfrac{\pi}{2}$, where even and odd values of the integer $k$ correspond to aster and vortex solutions, respectively.
Linear stability analysis detailed in~\hyperlink{appD}{Appendix D} shows that the former is always unstable, while the latter is stabilized by activity for a wide parameter range.
In particular, although reversals in Eq.~\eqref{eq_hydro_p} align the polarization radially,
the active coupling between polarity and order 
aligns the nematic order tangentially to the interface [Fig.~\ref{sketch_interactions}(c)]. 

To characterize the steady-state density profile associated with the vortex solution, 
we set $k$ odd and eliminate $\bS$ in~\eqref{ss_density} by solving the stationary Eq.~\eqref{eq_hydro_Q}.
Sufficiently far from the core, the magnitude of the nematic tensor is controlled by the Landau term (details in~\hyperlink{appC}{Appendix C}), giving
$\bar{S}(r) \approx \sqrt{2\mu(\bar{\rho}(r) - \rho_c)/\xi}$ for
$\bar{\rho}(r) > \rho_c$, and $\bar{S}(r) = 0$ otherwise.
Replacing the expressions for $\bar{S}$ and $\bar{\Phi}$ into Eq.~\eqref{ss_density} and assuming that $\bar{\rho}(r) \ge \rho_c$, we find
\begin{equation} \label{ss_density_closed}
    \left(1 - \frac{\varpi}{2\sqrt{\bar{\rho} - \rho_c}}
     \right)\bar{\rho}' - \frac{2\varpi}{ r}\sqrt{\bar{\rho} - \rho_c} + \zeta \bar{\rho} g(r) = 0,
\end{equation}
where $\varpi = \tilde{\pi}_2\tilde{D}_\rho^{-1}\sqrt{2\mu/\xi}$.
The density profile solving~\eqref{ss_density_closed} can be expressed in terms of Lambert functions, 
which approximate for thin interfaces and $r \lesssim R$ to
\begin{equation} \label{eq_rho_ring}
    \bar{\rho}(r) \underset{r^* \le r \lesssim R}{\simeq} \rho_c + \varpi^2 \ln^2\left( \frac{r}{r^*} \right), \quad
    r^* = R e^{-\tfrac{\sqrt{\rho_R - \rho_c}}{\varpi}},
\end{equation}
where $\rho_R$ denotes the peak density of the ring and $w=R-r^*= R(1 - \exp[-\sqrt{\rho_R - \rho_c}/\varpi])$
  its {\it extensive} width.
As we expect that $\rho_R \propto \rho_0$,
Eq.~\eqref{eq_rho_ring} predicts that $w$ also grows with $\rho_0$,
in qualitative agreement with the continuum and microscopic model simulations [Figs.~\ref{profiles}(a) and~\ref{hydro}(a)].
On the other hand, in the limit $r \to 0$ 
Eq.~\eqref{ss_density_closed} is well defined only when $\bar{\rho}(r
\to 0) \to \rho_c$ with a vanishing derivative, such that
$\bar{\rho}_{\rm core}(r \le r^*)=\rho_c$. 
We thus recover that the density at the centre of the vortex is exclusively determined by $\rho_c$
and independent of the global density $\rho_0$.
Finally, assuming that the nematic field is homogeneous outside of the illuminated disk,
the outside density profile is given by $\bar{\rho}_{\rm out}(r > R)
\approx \rho_R \exp(-\zeta)$, 
while $\rho_R$ is set by density normalization.
As shown in Fig.~\ref{hydro}(b), the piecewise solution built by combining $\bar{\rho}_{\rm core}$, $\bar{\rho}_{\rm out}$ and Eq.~\eqref{eq_rho_ring}
provides a reasonable approximation of the numerically obtained ring
profiles without fitting parameters. 
It also confirms the mechanism at the origin of
density protection: the density increase at $r \lesssim R$ is driven 
by the term $\propto\! \varpi^2\! \propto\! \tilde{\pi}_2^2$, 
originating from the curvature-induced active current ${\bf J} \! \propto \! {\bf p} \! \propto \! -\pi_2 \nabla \cdot {\bf Q}$.
Since this current vanishes when local nematic order is lost,
the core density must be determined by the threshold $\rho_c$.

Note also that the absence of microscopic repelling torques ($\beta
= 0$) makes the illuminated disk perfectly
trapping, corresponding to a diverging effective reversal rate $\kappa$.
In this limit, the active curvature-driven flux is suppressed and the ring state is lost,
consistent with both microscopic and continuum simulations (\hyperlink{appA}{Appendix A}).

To summarize, combining agent-based simulations with a minimal continuum description, 
we have uncovered a generic mechanism by which virtual confinement, induced by light-responsive activity, drives large-scale pattern formation beyond simple homogeneous accumulation. 
Collective effects endow the resulting structures with remarkable properties: 
although they originate from a localized driving, these patterns survive in the thermodynamic limit and exhibit density protection at macroscopic scales.
The robustness of this mechanism---which holds for general confining geometries and relies on minimal ingredients ubiquitous in active systems, namely alignment interactions and active currents---suggests that our results should be testable across a broad range of experimental platforms. 
Its applicability should moreover not be limited to light-responsive active matter,
as it may also be relevant for describing structure formation induced by more general types of confinement.
Finally, since Eq.~\eqref{eq_rho_ring} does not feature any intrinsic length scale, we further speculate that active currents similar to those generating density-protected states could be involved in the emergence of ring-like active filaments networks~\cite{NdlecNature1997,SuminoNature2012,SugiNatCom2019,FaluwekiPRL2023}, whose size is often orders of magnitude larger than that of individual filaments.\\

\begin{acknowledgments}
We are grateful to Hugues Chat\'e for a critical reading of the manuscript, 
and to Xiaqing Shi and Elia Bronzo for stimulating discussions.
G.F. and B.M. also acknowledge the hospitality of the Living Matter Department at MPI-DS, where part of this work was conducted.
G. F. and F. G. acknowledge support from grant PRIN 2020-PFCXPEAG from MIUR.
\end{acknowledgments}

\bibliography{biblio_abbr}

@article{CammannComPhys2024,
	author = {Cammann, Jan and Faluweki, Mixon K. and Dambacher, Nayara and Goehring, Lucas and Mazza, Marco G.},
	date = {2024/11/20},
	doi = {10.1038/s42005-024-01866-5},
	id = {Cammann2024},
	isbn = {2399-3650},
	journal = {Commun. Phys.},
	number = {1},
	pages = {376},
	title = {Topological transition in filamentous cyanobacteria: from motion to structure},
	url = {https://doi.org/10.1038/s42005-024-01866-5},
	volume = {7},
	year = {2024}
}

@misc{supplement,
	note = {{See Supplemental Material at [URL will be inserted by publisher] for further details}}}

@article{peshkov2012nonlinear,
	author = {Peshkov, Anton and Aranson, Igor S. and Bertin, Eric and Chat\'e, Hugues and Ginelli, Francesco},
	doi = {10.1103/PhysRevLett.109.268701},
	issue = {26},
	journal = {Phys. Rev. Lett.},
	month = {Dec},
	numpages = {5},
	pages = {268701},
	publisher = {American Physical Society},
	title = {Nonlinear Field Equations for Aligning Self-Propelled Rods},
	url = {https://link.aps.org/doi/10.1103/PhysRevLett.109.268701},
	volume = {109},
	year = {2012}}

@article{Ramaswamy,
	author = {Ramaswamy, Sriram},
	doi = {https://doi.org/10.1146/annurev-conmatphys-070909-104101},
	issn = {1947-5462},
	journal = {Annu. Rev. Condens. Matter Phys.},
	number = {Volume 1, 2010},
	pages = {323-345},
	publisher = {Annual Reviews},
	title = {The Mechanics and Statistics of Active Matter},
	type = {Journal Article},
	url = {https://www.annualreviews.org/content/journals/10.1146/annurev-conmatphys-070909-104101},
	volume = {1},
	year = {2010}}

@article{Romanzuck2024,
	author = {Romanczuk, Pawel and Chat{\'e}, Hugues and Chen, Leiming and Ngo, Sandrine and Toner, John},
	doi = {10.1088/1367-2630/18/6/063015},
	journal = {New J. Phys.},
	month = {jun},
	number = {6},
	pages = {063015},
	publisher = {IOP Publishing},
	title = {Emergent smectic order in simple active particle models},
	url = {https://doi.org/10.1088/1367-2630/18/6/063015},
	volume = {18},
	year = {2016}}

@article{Lenzini2024,
	author = {Lenzini, Leonardo and Fava, Giuseppe and Ginelli, Francesco},
	doi = {10.1088/1742-5468/ad6c2e},
	journal = {J. Stat. Mech.: Theory Exp.},
	month = {aug},
	number = {8},
	pages = {083210},
	publisher = {IOP Publishing},
	title = {Boundary symmetry breaking of flocking systems},
	url = {https://doi.org/10.1088/1742-5468/ad6c2e},
	volume = {2024},
	year = {2024}}

@article{ZhaoPRL2024,
	author = {Zhao, Zihui and Yao, Yisong and Li, He and Zhao, Yongfeng and Wang, Yujia and Zhang, Hepeng and Chat\'e, Hugues and Sano, Masaki},
	doi = {10.1103/PhysRevLett.133.268301},
	issue = {26},
	journal = {Phys. Rev. Lett.},
	month = {Dec},
	numpages = {8},
	pages = {268301},
	publisher = {American Physical Society},
	title = {Integer Topological Defects Reveal Antisymmetric Forces in Active Nematics},
	url = {https://link.aps.org/doi/10.1103/PhysRevLett.133.268301},
	volume = {133},
	year = {2024}}

@article{ZhaoNatCom2025,
	author = {Zhao, Zihui and Li, He and Yao, Yisong and Zhao, Yongfeng and Serra, Francesca and Kawaguchi, Kyogo and Zhang, Hepeng and Sano, Masaki},
	date = {2025/03/12},
	doi = {10.1038/s41467-025-57783-w},
	id = {Zhao2025},
	isbn = {2041-1723},
	journal = {Nat. Commun.},
	number = {1},
	pages = {2452},
	title = {Integer topological defects offer a methodology to quantify and classify active cell monolayers},
	url = {https://doi.org/10.1038/s41467-025-57783-w},
	volume = {16},
	year = {2025}}

@article{KaiyrbekovPNAS2023,
	author = {Kurmanbek Kaiyrbekov and Kirsten Endresen and Kyle Sullivan and Zhaofei Zheng and Yun Chen and Francesca Serra and Brian A. Camley},
	doi = {10.1073/pnas.2301197120},
	journal = {Proc. Natl. Acad. Sci. U.S.A.},
	number = {30},
	pages = {e2301197120},
	title = {Migration and division in cell monolayers on substrates with topological defects},
	url = {https://www.pnas.org/doi/abs/10.1073/pnas.2301197120},
	volume = {120},
	year = {2023}}

@article{RisserAnnrev2025,
	author = {Risser, Douglas D.},
	doi = {https://doi.org/10.1146/annurev-micro-051024-033328},
	issn = {1545-3251},
	journal = {Annu. Rev. Microbiol.},
	number = {Volume 79, 2025},
	pages = {69-85},
	publisher = {Annual Reviews},
	title = {Motility in Filamentous Cyanobacteria},
	type = {Journal Article},
	url = {https://www.annualreviews.org/content/journals/10.1146/annurev-micro-051024-033328},
	volume = {79},
	year = {2025}}

@article{RamaswamyEPL2003,
	author = {S. Ramaswamy and R. Aditi Simha and J. Toner},
	doi = {10.1209/epl/i2003-00346-7},
	journal = {EPL},
	month = {apr},
	number = {2},
	pages = {196},
	title = {Active nematics on a substrate: Giant number fluctuations and long-time tails},
	url = {https://doi.org/10.1209/epl/i2003-00346-7},
	volume = {62},
	year = {2003}}

@article{MahaultJSTAT2018,
	author = {Mahault, Beno{\^\i}t and Patelli, Aurelio and Chat\'e, Hugues},
	doi = {10.1088/1742-5468/aad6b5},
	journal = {J. Stat. Mech.: Theory Exp.},
	month = {sep},
	number = {9},
	pages = {093202},
	publisher = {IOP Publishing and SISSA},
	title = {Deriving hydrodynamic equations from dry active matter models in three dimensions},
	url = {https://doi.org/10.1088/1742-5468/aad6b5},
	volume = {2018},
	year = {2018}}

@article{BertinPRE2015,
	author = {Bertin, Eric and Baskaran, Aparna and Chat\'e, Hugues and Marchetti, M. Cristina},
	doi = {10.1103/PhysRevE.92.042141},
	issue = {4},
	journal = {Phys. Rev. E},
	month = {Oct},
	numpages = {10},
	pages = {042141},
	publisher = {American Physical Society},
	title = {Comparison between Smoluchowski and Boltzmann approaches for self-propelled rods},
	url = {https://link.aps.org/doi/10.1103/PhysRevE.92.042141},
	volume = {92},
	year = {2015}}

@book{DeGennesProst,
	author = {De Gennes, P G and Prost, J},
	doi = {10.1093/oso/9780198520245.001.0001},
	isbn = {9780198520245},
	month = {12},
	publisher = {Oxford University Press},
	title = {The Physics of Liquid Crystals},
	url = {https://doi.org/10.1093/oso/9780198520245.001.0001},
	year = {1993}}

@article{NdlecNature1997,
	author = {Ndlec, F. J. and Surrey, T. and Maggs, A. C. and Leibler, S.},
	date = {1997/09/01},
	doi = {10.1038/38532},
	id = {Ndlec1997},
	isbn = {1476-4687},
	journal = {Nature},
	number = {6648},
	pages = {305--308},
	title = {Self-organization of microtubules and motors},
	url = {https://doi.org/10.1038/38532},
	volume = {389},
	year = {1997}}

@article{WinklerJCP2020,
	author = {Winkler, Roland G. and Gompper, Gerhard},
	doi = {10.1063/5.0011466},
	issn = {0021-9606},
	journal = {J. Chem. Phys.},
	month = {07},
	number = {4},
	pages = {040901},
	title = {The physics of active polymers and filaments},
	url = {https://doi.org/10.1063/5.0011466},
	volume = {153},
	year = {2020}}

@article{MijalkovPRX2016,
	author = {Mijalkov, Mite and McDaniel, Austin and Wehr, Jan and Volpe, Giovanni},
	doi = {10.1103/PhysRevX.6.011008},
	issue = {1},
	journal = {Phys. Rev. X},
	month = {Jan},
	numpages = {16},
	pages = {011008},
	publisher = {American Physical Society},
	title = {Engineering Sensorial Delay to Control Phototaxis and Emergent Collective Behaviors},
	url = {https://link.aps.org/doi/10.1103/PhysRevX.6.011008},
	volume = {6},
	year = {2016}}

@article{ShankarPNAS2024,
	author = {Suraj Shankar and Luca V. D. Scharrer and Mark J. Bowick and M. Cristina Marchetti},
	doi = {10.1073/pnas.2400933121},
	journal = {Proc. Natl. Acad. Sci. U.S.A.},
	number = {21},
	pages = {e2400933121},
	title = {Design rules for controlling active topological defects},
	url = {https://www.pnas.org/doi/abs/10.1073/pnas.2400933121},
	volume = {121},
	year = {2024}}

@article{BenDorPRE2022,
	author = {Ben Dor, Ydan and Ro, Sunghan and Kafri, Yariv and Kardar, Mehran and Tailleur, Julien},
	doi = {10.1103/PhysRevE.105.044603},
	issue = {4},
	journal = {Phys. Rev. E},
	month = {Apr},
	numpages = {16},
	pages = {044603},
	publisher = {American Physical Society},
	title = {Disordered boundaries destroy bulk phase separation in scalar active matter},
	url = {https://link.aps.org/doi/10.1103/PhysRevE.105.044603},
	volume = {105},
	year = {2022}}

@article{CodinaPRL2022,
	author = {Codina, Joan and Mahault, Beno\^{\i}t and Chat{\'e}, Hugues and Dobnikar, Jure and Pagonabarraga, Ignacio and Shi, Xia-qing},
	doi = {10.1103/PhysRevLett.128.218001},
	issue = {21},
	journal = {Phys. Rev. Lett.},
	month = {May},
	numpages = {6},
	pages = {218001},
	publisher = {American Physical Society},
	title = {Small Obstacle in a Large Polar Flock},
	url = {https://link.aps.org/doi/10.1103/PhysRevLett.128.218001},
	volume = {128},
	year = {2022}}

@article{BowickPRX2022,
	author = {Bowick, Mark J. and Fakhri, Nikta and Marchetti, M. Cristina and Ramaswamy, Sriram},
	doi = {10.1103/PhysRevX.12.010501},
	issue = {1},
	journal = {Phys. Rev. X},
	month = {Feb},
	numpages = {27},
	pages = {010501},
	publisher = {American Physical Society},
	title = {Symmetry, Thermodynamics, and Topology in Active Matter},
	url = {https://link.aps.org/doi/10.1103/PhysRevX.12.010501},
	volume = {12},
	year = {2022}}

@article{GinelliPRL2010,
	author = {Ginelli, Francesco and Peruani, Fernando and B\"ar, Markus and Chat{\'e}, Hugues},
	doi = {10.1103/PhysRevLett.104.184502},
	issue = {18},
	journal = {Phys. Rev. Lett.},
	month = {May},
	numpages = {4},
	pages = {184502},
	publisher = {American Physical Society},
	title = {{Large-Scale Collective Properties of Self-Propelled Rods}},
	url = {https://link.aps.org/doi/10.1103/PhysRevLett.104.184502},
	volume = {104},
	year = {2010}}

@article{LozanoNatCom2016,
	author = {Lozano, Celia and ten Hagen, Borge and L{\"o}wen, Hartmut and Bechinger, Clemens},
	date = {2016/09/30},
	doi = {10.1038/ncomms12828},
	id = {Lozano2016},
	isbn = {2041-1723},
	journal = {Nat. Commun.},
	number = {1},
	pages = {12828},
	title = {Phototaxis of synthetic microswimmers in optical landscapes},
	url = {https://doi.org/10.1038/ncomms12828},
	volume = {7},
	year = {2016}}

@article{AlvarezNatCom2021,
	author = {Alvarez, L. and Fernandez-Rodriguez, M. A. and Alegria, A. and Arrese-Igor, S. and Zhao, K. and Kr{\"o}ger, M. and Isa, Lucio},
	date = {2021/08/06},
	doi = {10.1038/s41467-021-25108-2},
	id = {Alvarez2021},
	isbn = {2041-1723},
	journal = {Nat. Commun.},
	number = {1},
	pages = {4762},
	title = {Reconfigurable artificial microswimmers with internal feedback},
	url = {https://doi.org/10.1038/s41467-021-25108-2},
	volume = {12},
	year = {2021}}

@article{vanKesterenPNAS2023,
	author = {Steven van Kesteren and Laura Alvarez and Silvia Arrese-Igor and Angel Alegria and Lucio Isa},
	doi = {10.1073/pnas.2213481120},
	journal = {Proc. Natl. Acad. Sci. U.S.A.},
	number = {11},
	pages = {e2213481120},
	title = {Self-propelling colloids with finite state dynamics},
	url = {https://www.pnas.org/doi/abs/10.1073/pnas.2213481120},
	volume = {120},
	year = {2023}}

@article{SokerPRL2021,
	author = {S\"oker, Nicola Andreas and Auschra, Sven and Holubec, Viktor and Kroy, Klaus and Cichos, Frank},
	doi = {10.1103/PhysRevLett.126.228001},
	issue = {22},
	journal = {Phys. Rev. Lett.},
	month = {Jun},
	numpages = {6},
	pages = {228001},
	publisher = {American Physical Society},
	title = {How Activity Landscapes Polarize Microswimmers without Alignment Forces},
	url = {https://link.aps.org/doi/10.1103/PhysRevLett.126.228001},
	volume = {126},
	year = {2021}}

@article{FrangipaneeLife2018,
	article_type = {journal},
	author = {Frangipane, Giacomo and Dell'Arciprete, Dario and Petracchini, Serena and Maggi, Claudio and Saglimbeni, Filippo and Bianchi, Silvio and Vizsnyiczai, Gaszton and Bernardini, Maria Lina and Di Leonardo, Roberto},
	citation = {eLife 2018;7:e36608},
	doi = {10.7554/eLife.36608},
	editor = {Goldstein, Raymond E and Barkai, Naama},
	issn = {2050-084X},
	journal = {eLife},
	month = {aug},
	pages = {e36608},
	pub_date = {2018-08-14},
	publisher = {eLife Sciences Publications, Ltd},
	title = {Dynamic density shaping of photokinetic \textit{E. coli}},
	url = {https://doi.org/10.7554/eLife.36608},
	volume = 7,
	year = 2018}

@article{Massana-CidNatCom2022,
	author = {Massana-Cid, Helena and Maggi, Claudio and Frangipane, Giacomo and Di Leonardo, Roberto},
	date = {2022/05/18},
	doi = {10.1038/s41467-022-30201-1},
	id = {Massana-Cid2022},
	isbn = {2041-1723},
	journal = {Nat. Commun.},
	number = {1},
	pages = {2740},
	title = {Rectification and confinement of photokinetic bacteria in an optical feedback loop},
	url = {https://doi.org/10.1038/s41467-022-30201-1},
	volume = {13},
	year = {2022}}

@article{ArltNatCom2018,
	author = {Arlt, Jochen and Martinez, Vincent A. and Dawson, Angela and Pilizota, Teuta and Poon, Wilson C. K.},
	date = {2018/02/22},
	doi = {10.1038/s41467-018-03161-8},
	id = {Arlt2018},
	isbn = {2041-1723},
	journal = {Nat. Commun.},
	number = {1},
	pages = {768},
	title = {Painting with light-powered bacteria},
	url = {https://doi.org/10.1038/s41467-018-03161-8},
	volume = {9},
	year = {2018}}

@article{WildeFEMS2017,
	author = {Wilde, Annegret and Mullineaux, Conrad W.},
	doi = {10.1093/femsre/fux045},
	issn = {0168-6445},
	journal = {FEMS Microbiol. Rev.},
	month = {10},
	number = {6},
	pages = {900-922},
	title = {Light-controlled motility in prokaryotes and the problem of directional light perception},
	url = {https://doi.org/10.1093/femsre/fux045},
	volume = {41},
	year = {2017}}

@article{SugiNatCom2019,
	author = {Sugi, Takuma and Ito, Hiroshi and Nishimura, Masaki and Nagai, Ken H.},
	date = {2019/02/18},
	doi = {10.1038/s41467-019-08537-y},
	id = {Sugi2019},
	isbn = {2041-1723},
	journal = {Nat. Commun.},
	number = {1},
	pages = {683},
	title = {{C. elegans} collectively forms dynamical networks},
	url = {https://doi.org/10.1038/s41467-019-08537-y},
	volume = {10},
	year = {2019}}

@article{FaluwekiPRL2023,
	author = {Faluweki, Mixon K. and Cammann, Jan and Mazza, Marco G. and Goehring, Lucas},
	doi = {10.1103/PhysRevLett.131.158303},
	issue = {15},
	journal = {Phys. Rev. Lett.},
	month = {Oct},
	numpages = {7},
	pages = {158303},
	publisher = {American Physical Society},
	title = {Active Spaghetti: Collective Organization in Cyanobacteria},
	url = {https://link.aps.org/doi/10.1103/PhysRevLett.131.158303},
	volume = {131},
	year = {2023}}

@article{SuminoNature2012,
	author = {Sumino, Yutaka and Nagai, Ken H. and Shitaka, Yuji and Tanaka, Dan and Yoshikawa, Kenichi and Chat{\'e}, Hugues and Oiwa, Kazuhiro},
	date = {2012/03/01},
	doi = {10.1038/nature10874},
	id = {Sumino2012},
	isbn = {1476-4687},
	journal = {Nature},
	number = {7390},
	pages = {448--452},
	title = {Large-scale vortex lattice emerging from collectively moving microtubules},
	url = {https://doi.org/10.1038/nature10874},
	volume = {483},
	year = {2012}}

@article{SchallerNature2010,
	author = {Schaller, Volker and Weber, Christoph and Semmrich, Christine and Frey, Erwin and Bausch, Andreas R.},
	date = {2010/09/01},
	doi = {10.1038/nature09312},
	id = {Schaller2010},
	isbn = {1476-4687},
	journal = {Nature},
	number = {7311},
	pages = {73--77},
	title = {Polar patterns of driven filaments},
	url = {https://doi.org/10.1038/nature09312},
	volume = {467},
	year = {2010}}

@article{DavisPRX2024,
	author = {Davis, Luke K. and Proesmans, Karel and Fodor, \'Etienne},
	doi = {10.1103/PhysRevX.14.011012},
	issue = {1},
	journal = {Phys. Rev. X},
	month = {Feb},
	numpages = {19},
	pages = {011012},
	publisher = {American Physical Society},
	title = {Active Matter under Control: Insights from Response Theory},
	url = {https://link.aps.org/doi/10.1103/PhysRevX.14.011012},
	volume = {14},
	year = {2024}}

@article{NortonPRL2020,
	author = {Norton, Michael M. and Grover, Piyush and Hagan, Michael F. and Fraden, Seth},
	doi = {10.1103/PhysRevLett.125.178005},
	issue = {17},
	journal = {Phys. Rev. Lett.},
	month = {Oct},
	numpages = {7},
	pages = {178005},
	publisher = {American Physical Society},
	title = {Optimal Control of Active Nematics},
	url = {https://link.aps.org/doi/10.1103/PhysRevLett.125.178005},
	volume = {125},
	year = {2020}}

@article{SchupplerNatCom2016,
	author = {Schuppler, Matthias and Keber, Felix C. and Kr{\"o}ger, Martin and Bausch, Andreas R.},
	date = {2016/10/14},
	doi = {10.1038/ncomms13120},
	id = {Schuppler2016},
	isbn = {2041-1723},
	journal = {Nat. Commun.},
	number = {1},
	pages = {13120},
	title = {Boundaries steer the contraction of active gels},
	url = {https://doi.org/10.1038/ncomms13120},
	volume = {7},
	year = {2016}}

@article{ZhangNatMat2021,
	author = {Zhang, Rui and Redford, Steven A. and Ruijgrok, Paul V. and Kumar, Nitin and Mozaffari, Ali and Zemsky, Sasha and Dinner, Aaron R. and Vitelli, Vincenzo and Bryant, Zev and Gardel, Margaret L. and de Pablo, Juan J.},
	date = {2021/06/01},
	doi = {10.1038/s41563-020-00901-4},
	id = {Zhang2021},
	isbn = {1476-4660},
	journal = {Nat. Mater.},
	number = {6},
	pages = {875--882},
	title = {Spatiotemporal control of liquid crystal structure and dynamics through activity patterning},
	url = {https://doi.org/10.1038/s41563-020-00901-4},
	volume = {20},
	year = {2021}}

@article{ShankarNRP2022,
	author = {Shankar, Suraj and Souslov, Anton and Bowick, Mark J. and Marchetti, M. Cristina and Vitelli, Vincenzo},
	date = {2022/06/01},
	doi = {10.1038/s42254-022-00445-3},
	id = {Shankar2022},
	isbn = {2522-5820},
	journal = {Nat. Rev. Phys.},
	number = {6},
	pages = {380--398},
	title = {Topological active matter},
	url = {https://doi.org/10.1038/s42254-022-00445-3},
	volume = {4},
	year = {2022}}

@article{LiuNature2021,
	author = {Liu, Song and Shankar, Suraj and Marchetti, M. Cristina and Wu, Yilin},
	date = {2021/02/01},
	doi = {10.1038/s41586-020-03168-6},
	id = {Liu2021},
	isbn = {1476-4687},
	journal = {Nature},
	number = {7844},
	pages = {80--84},
	title = {Viscoelastic control of spatiotemporal order in bacterial active matter},
	url = {https://doi.org/10.1038/s41586-020-03168-6},
	volume = {590},
	year = {2021}}

@article{PatelliPRL2019,
	author = {Patelli, Aurelio and Djafer-Cherif, Ilyas and Aranson, Igor S. and Bertin, Eric and Chat{\'e}, Hugues},
	doi = {10.1103/PhysRevLett.123.258001},
	issue = {25},
	journal = {Phys. Rev. Lett.},
	month = {Dec},
	numpages = {6},
	pages = {258001},
	publisher = {American Physical Society},
	title = {Understanding Dense Active Nematics from Microscopic Models},
	url = {https://link.aps.org/doi/10.1103/PhysRevLett.123.258001},
	volume = {123},
	year = {2019}}

@article{KurjahnNC2024,
	author = {Kurjahn, Maximilian and Abbaspour, Leila and Papenfu{\ss}, Franziska and Bittihn, Philip and Golestanian, Ramin and Mahault, Beno{\^\i}t and Karpitschka, Stefan},
	date = {2024/10/23},
	doi = {10.1038/s41467-024-52936-9},
	id = {Kurjahn2024},
	isbn = {2041-1723},
	journal = {Nat. Commun.},
	number = {1},
	pages = {9122},
	title = {Collective self-caging of active filaments in virtual confinement},
	url = {https://doi.org/10.1038/s41467-024-52936-9},
	volume = {15},
	year = {2024}}

@article{RossNature2019,
	author = {Ross, Tyler D. and Lee, Heun Jin and Qu, Zijie and Banks, Rachel A. and Phillips, Rob and Thomson, Matt},
	date = {2019/08/01},
	doi = {10.1038/s41586-019-1447-1},
	id = {Ross2019},
	isbn = {1476-4687},
	journal = {Nature},
	number = {7768},
	pages = {224--229},
	title = {Controlling organization and forces in active matter through optically defined boundaries},
	url = {https://doi.org/10.1038/s41586-019-1447-1},
	volume = {572},
	year = {2019}}

@article{Fava2024PRL,
	author = {Fava, Giuseppe and Gambassi, Andrea and Ginelli, Francesco},
	doi = {10.1103/PhysRevLett.133.148301},
	issue = {14},
	journal = {Phys. Rev. Lett.},
	month = {Oct},
	numpages = {6},
	pages = {148301},
	publisher = {American Physical Society},
	title = {{Strong Casimir-like Forces in Flocking Active Matter}},
	url = {https://link.aps.org/doi/10.1103/PhysRevLett.133.148301},
	volume = {133},
	year = {2024}}

@article{DADAM,
	author = {Chat\'e, Hugues},
	journal = {Annu. Rev. Condens. Matter Phys.},
	number = {11},
	pages = {182},
	title = {Dry Aligning Dilute Active Matter},
	year = {2020}}

\onecolumngrid

\newpage

\vspace{12pt}
\noindent\hrulefill \hspace{24pt} {\bf End Matter} \hspace{24pt} \hrulefill
\vspace{12pt}

\twocolumngrid

\renewcommand \thefigure{A\arabic{figure}}
\setcounter{figure}{0}
\setcounter{equation}{0}

\hypertarget{appA}{\paragraph{Appendix A: The dynamics without repelling torques---}}
For parameters of the microscopic model considered in the main text, 
rings form for a small but finite value ($\approx 10^{-2}$) of $\beta$.
Since the repelling torque strength is an effective parameter not
controllable experimentally, this transition will be investigated
elsewhere.

At the hydrodynamic level, we argue in the main text that the absence of microscopic repelling torques corresponds to the regime $\kappa \to \infty$.
In this limit, the polarity $\bar\p$ is simply enslaved to the external field $\kappa g(r) \rho \n(\varphi)$
so that $\pi_0, \, \pi_2\approx 0$, suppressing the active flux responsible for density accumulation at the disk boundary. 
Numerical simulations of Eqs.~\eqref{eq_hydro} with $\pi_2 = 0$ indeed reveal that in this regime
the density profile remains nearly homogeneous inside and outside the illuminated disk (not shown),
in agreement with Eq.~\eqref{eq_rho_noring} and the analysis carried out in~\hyperlink{appC}{Appendix C}.
Repelling torques are thus
essential for ring formation.\\

\setcounter{figure}{0}
\renewcommand \thefigure{B\arabic{figure}}
\hypertarget{appB}{\paragraph{Appendix B: Density protection in noncircular geometries---}} In addition to the circular geometry discussed in the main text, Fig.~\ref{noncircular_accumulation} demonstrates that density protection persists in more general convex and nonconvex domains. 
Specifically, we simulate Eqs.~\eqref{micro} imposing two different confining shapes: a square of side length $2R = 180$ [Figs.~\ref{noncircular_accumulation}(a,c)] and a superellipse with semi-diameter $R = 96$ and parameter $n = 1/2$ [Figs.~\ref{noncircular_accumulation}(b,d)].
In both geometries, particles accumulate near and align with the virtual boundary when $\beta > 0$ and the mean particle density $\rho_0$ is sufficiently high,
while the density at the centre of the illuminated domain remains independent of $\rho_0$.\\

%%%%%%%%%%%%%%%%%%%%%%%%%%%%%%
\begin{figure}[t!]
\includegraphics[width=\linewidth]{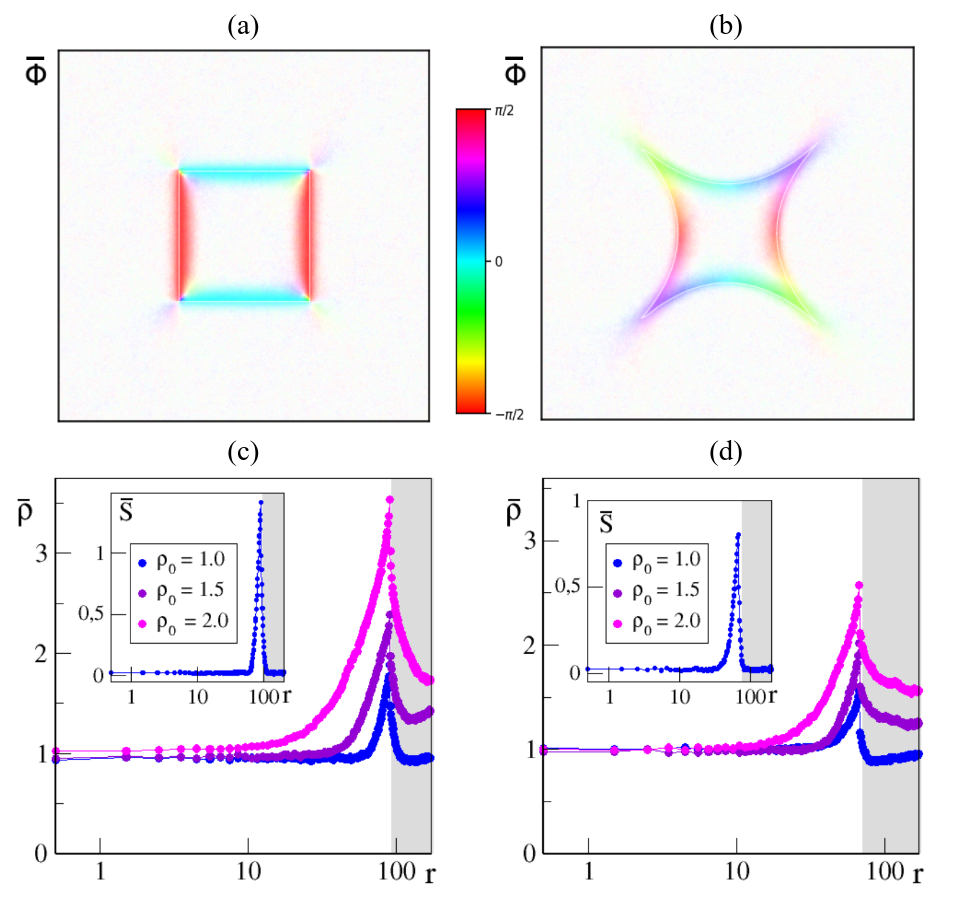}
\caption{Boundary accumulation in square (a,c) and superellipse (b,d) geometries.
Panels (a,b) show the (color-coded) time-averaged nematic angle field, where disordered regions appear white and $\rho_0 = 1$.
(c,d) display the stationary density profiles averaged over the vertical and horizontal axes from the centre of the domain for several values of $\rho_0$. Insets: corresponding nematic amplitudes for $\rho_0 = 1$.
Parameters: $\beta = 0.6$ and $L = 512$.}
\label{noncircular_accumulation}
\end{figure}
%%%%%%%%%%%%%%%%%%%%%%%%%%%%%%

\setcounter{equation}{0}
\renewcommand \theequation{C\arabic{equation}}
\hypertarget{appC}{\paragraph{Appendix C: The steady-state solutions of Eqs.~\eqref{eq_hydro}---}}
As detailed in the main text, Eq.~\eqref{ss_density} admits three types of solution termed respectively homogeneous, aster and vortex.
The homogeneous solution, which is valid only in the infinitely thin interface limit where $g(r) = \delta(r - R)$, leads to uniform density profiles in and out of the illuminated disk [Eq.~\eqref{eq_rho_noring}]. 
The nematic amplitude is then also homogeneous in both regions and is found by solving $\left[\mu(\brho - \rho_c) - \xi'\bS^2\right]\bS = 0$,
while the nematic orientation $\bPhi$ is uniform across the whole system.

Aster and vortex solutions are characterized by an orientation profile $\bPhi(\varphi) = \varphi + k \tfrac{\pi}{2}$, where $k$ is an integer. 
With this ansatz, the stationary density and nematic amplitude solve
\begin{subequations} \label{eq_vortex}
\begin{align} \label{eq_rho_vortex}
    & \tilde{D}_\rho \brho' + \tilde{\pi}_2 (-1)^k \left(\bS' + \frac{2 \bS}{r}\right) + \frac{\kappa v_0}{\alpha} \brho g = 0, \\
    & \left[\mu(\brho - \rho_c) - \frac{\xi}{2}\bS^2\right]\bS + D_Q \left[\bS'' + \frac{1}{r}\bS' - \frac{4}{r^2}\bS\right] \nonumber \\
    \label{eq_S_vortex}
    & - \frac{\pi_1 D_\rho}{v_0}(-1)^k\left(\brho'' - \frac{1}{r}\brho'\right) = 0.
\end{align}
\end{subequations}
For simplicity, we now consider the thin interface limit and set $g(r) = 0$ for both $r < R$ and $r > R$. 

As the nematic order exhibits a defect at $r = 0$, solutions of Eqs.~\eqref{eq_vortex} are dominated by different contributions close to and far from the origin.
In the limit $r \to 0$, Eqs.~\eqref{eq_vortex} admit solutions of the form
\begin{equation} \label{sol_smallr}
\brho(r) \simeq \brho(0) + A r^2, \quad \bS(r) \simeq B r^2,
\end{equation} 
where $A$ and $B$ are two undetermined constants that satisfy
$A = (-1)^{k+1} 2 \tilde{\pi}_2 B / \tilde{D}_\rho$.
When local nematic order is present near the centre of the domain, $B > 0$ and aster configurations ($k$ even) lead to $A < 0$, such that density is highest at $r = 0$. 
Conversely, vortices with odd $k$ give $A > 0$, 
such that particles are advected towards the boundary of the disk.

We now consider the opposite limit of $r \lesssim R$. 
Since far from the centre of the disk derivatives of $\brho$ and $\bS$ scale like $r^{-1}$, 
for $R$ large the leading order term of Eq.~\eqref{eq_S_vortex} is now the Landau potential, leading to 
$\bS(r) \simeq \sqrt{2\mu[\brho(r) - \rho_c]/\xi}$ for $\brho(r) \ge \rho_c$ and $0$ otherwise.
Replacing this expression in Eq.~\eqref{eq_rho_vortex} and assuming that $\brho(r) \ge \rho_c$, 
we obtain
\begin{equation} \label{eq_rho_radial2}
    \left(1 + (-1)^k\frac{\varpi}{2\sqrt{\brho - \rho_c}}
     \right)\brho' + (-1)^k\frac{2\varpi}{ r}\sqrt{\brho - \rho_c} = 0,
\end{equation}
where $\varpi = \tilde{\pi}_2\tilde{D}_\rho^{-1}\sqrt{2\mu/\xi}$.
Solutions of Eq.~\eqref{eq_rho_radial2} can be written as $\brho(r) = \rho_c + X^2(r)$,
where $X(r)$ is expressed in terms of Lambert $W$ functions:
\begin{equation} \label{sol_larger}
    X(r) = (-1)^{k}\frac{\varpi}{2} W_n\left[ \frac{(-1)^{k} 2 C}{\varpi r^2} \right],
\end{equation}
where $n = 0$ or $-1$ according to the relevant branch, and $C$ is an integration constant.
For $k$ even, the argument of the Lambert function is positive, which implies that $n = 0$. 
Since $W_0$ is monotonously growing, $\brho(r)$ decreases as $r$ grows,
confirming that density is maximum at $r = 0$.

For $k$ odd, the argument and prefactor of $W_n$ are negative.
Since the density can take arbitrary large values, we must choose $n = -1$ 
as $W_{-1}(z)$ varies in $[-1;-\infty)$ for $z < 0$.
Using the identity $W_{-1}(x e^x) = x$ for $x \le -1$, 
we fix $C$ with the value $\rho_R$ of the density at $r = R$:
\begin{equation} \label{eq_X_sol}
    X(r) %= -\frac{\varpi}{2} W_{-1}\left[ -\frac{2\sqrt{\rho_R - \rho_c}}{\varpi} e^{-\tfrac{2\sqrt{\rho_R - \rho_c}}{\varpi}} \frac{R^2}{r^2} \right] 
    = -\frac{\varpi}{2} W_{-1}\left( -\frac{1}{e} \frac{{R^*}^2}{r^2} \right)
    \qquad (k \, {\rm odd}),
\end{equation}
where 
$
R^* = R \exp(\tfrac{1}{2} - \sqrt{\rho_R - \rho_c}/\varpi)(2\sqrt{\rho_R - \rho_c}/\varpi)^{1/2}
$.
Since the function $W_{-1}(z)$ is defined in the finite interval $[-e^{-1};0)$,
the solution~\eqref{eq_X_sol} only formally holds for $R^* \le r \le R$.
For $r < R^*$, on the other hand, the slowly varying fields assumption breaks down and 
the solution~\eqref{eq_X_sol} is no more defined.
In the limit where $\sqrt{\rho_R - \rho_c}/\varpi$ is large, however, 
$R^* \to 0$ and the asymptotic series $W_{-1}(z) \simeq \ln |z|$ for $z \to 0$
allows to recover Eq.~\eqref{eq_rho_ring},
which can be matched with Eq.~\eqref{sol_smallr} with $A=B=0$ for $r<r^*$.\\

\setcounter{equation}{0}
\renewcommand \theequation{D\arabic{equation}}
\hypertarget{appD}{\paragraph{Appendix D: Linear stability analysis of inhomogeneous solutions---}}
Here, we study the stability of the vortex and aster solutions of Eqs.~\eqref{eq_hydro} to perturbations of the form
$\rho = \brho(r) + \delta \rho(\varphi,t)$, $S = \bS(r) + \delta S(\varphi,t)$ 
and $\Phi = \bar\Phi(\varphi) + \delta\Phi(\varphi,t)$.
Going into angular Fourier space $(\delta \rho,\delta S,\delta\Phi) = \frac{1}{2\pi}\sum_n (\delta \rho_n,\delta S_n,\delta\Phi_n) e^{-i n \varphi}$,
the perturbations satisfy at linear order
\begin{subequations} \label{eq_noint_hydro_enslaved_exp}
    \begin{align} 
        \label{eq_noint_hydro_enslaved_exp_rho}
        \delta\dot{\rho}_n & = -\frac{\tilde{D}_\rho n^2}{r^2}\delta\rho_n
        + \Gamma_{\rho S} \delta S_n  + \Gamma_{\rho\Phi} \delta\Phi_n, \\
        \label{eq_noint_hydro_enslaved_exp_S}
        \delta \dot{S}_n & = \Gamma_{S\rho} \delta\rho_n 
        - {a}_n^2 \delta S_n 
        + \Gamma_{S\Phi}  \delta \Phi_n, \\
        \label{eq_noint_hydro_enslaved_exp_phi}
        \delta\dot{\Phi}_n & = \Gamma_{\Phi\rho} \delta\rho_n
        + \Gamma_{\Phi S}\delta S_n
        - K_n \delta\Phi_n,
    \end{align}
\end{subequations}
where 
\begin{align*}
\Gamma_{\rho S} & = \frac{(-1)^k \tilde{\pi}_2 n^2}{r^2}, 
& \Gamma_{\rho\Phi} & = - 4 i n \frac{(-1)^k \tilde{\pi}_2 (r \bS)' }{r^2}, \\
\Gamma_{S\rho} & = \mu \bS + \frac{(-1)^k\tilde{\pi}_0 n^2}{r^2},
& \Gamma_{S\Phi} & = 8 i n \frac{\tilde{D}_Q \bS}{r^2},\\
\Gamma_{\Phi\rho} & = in \frac{(-1)^k\tilde{\pi}_0}{r^2 \bS}, 
& \Gamma_{\Phi S} & = - 2 i n \frac{\tilde{D}_Q}{\bS r^2},
\end{align*}
while
%\begin{align*}
$a_n^2 = -\mu(\brho - \rho_c) + \tfrac{3}{2}\xi \bS^2 + r^{-2}\tilde{D}_Q(n^2 + 4)$,
$$
K_n = \frac{(-1)^k \tilde{\pi}_0}{\bS} r \left(\frac{\brho'}{r}\right)' + \frac{\tilde{D}_Q n^2}{r^2},
$$
$\tilde{\pi}_0 = \pi_1 \pi_0/\alpha$,
and $\tilde D_Q = D_Q + \tfrac{1}{2}\pi_1\pi_2/\alpha$.

We first consider $n = 0$ perturbations. 
While $a_0^2$ generally remains positive for both aster and vortex solutions, 
we deduce from Eqs.~\eqref{eq_rho_ring} and~\eqref{sol_smallr} that 
$K_0$ takes the following values in regions where particles accumulate 
\begin{equation} \label{eqK}
    K_0 \underset{r \to 0}{\simeq} 0 \; ({\rm aster}), 
    \qquad 
    K_0 \underset{r \lesssim R}{\simeq} \frac{4 \tilde{\pi}_0\tilde{\pi}_2}{\tilde{D}_\rho r^2} \quad ({\rm vortex}).
\end{equation}
Close to the centre of the disk, the aster solution is thus only marginally stable to $n = 0$ perturbations.
Hence, in the presence of fluctuations this solution should rapidly be destroyed by rotations of the nematic order. 
For the ring solution, on the other hand, $\bS$ is exactly $0$ at small $r$ while $K_0 > 0$ for large $r < R$.
In addition, we note from~\eqref{eqK} that $K(r) \propto \pi_0 \pi_2$, indicating that the vortex structure is stabilized by activity.

The fate of finite $n$ perturbations is determined by the characteristic polynomial of the linear system~\eqref{eq_noint_hydro_enslaved_exp},
which is discussed in details in the Supplemental Material~\cite{supplement}.
The analysis presented there allows to conclude that rings must be sufficiently large and dense to be stable to linear perturbations.
In addition, systems for which the effective parameter $\varpi$ is small tend to be more prone to exhibit stable ring structures.\\

\hypertarget{appE}{\paragraph{Appendix E: Details on numerical simulations.---}}
Simulations of Eqs.~\eqref{eq_hydro} were performed by means of a pseudo-spectral solver with $1/3$ de-aliasing and an Euler explicit scheme for time integration. For all simulations, we described the interface with the function $g(r) = \exp( - (r - R)^2 / 2\sigma^2 ) / \sqrt{2 \pi \sigma^2}$ and $\sigma = 1/4$. Space and time resolutions were taken equal to $dx = 1/4$ and $dt = 0.01$, respectively.

\clearpage

\end{document}